\documentclass{appolb}
\usepackage{graphicx}
\usepackage{color}

\def\be{\begin{equation}}
\def\ee{\end{equation}}
\def\bea{\begin{eqnarray}}
\def\eea{\end{eqnarray}}
\newcommand{\m}{M_{H^{\pm}}}
\newcommand{\rg}{R_{\gamma\gamma}}
\usepackage{amsmath}
\usepackage{amssymb}
\usepackage{amsthm}
\usepackage{amsfonts}

\newcommand{\rgg}{\mathcal{R}_{\gamma\gamma}}
 
\newcommand{\la}{\lambda_1}
\newcommand{\lb}{\lambda_2}
\newcommand{\lc}{\lambda_3}
\newcommand{\lczp}{\lambda_{345}}

\newcommand{\g}{{\rm {GeV}}}

\begin{document}
\title{The Inert Doublet Model and its extensions%
\thanks{Presented at Zakopane School 2015 by M. Krawczyk and N. Darvishi.}%
}
\author{Maria Krawczyk, Neda Darvishi, Dorota Soko\l owska
\address{Faculty of Physics, University of Warsaw \\ ul. Pasteura 5, 02-093 Warsaw, Poland}
\\
}
\maketitle
\begin{abstract}
The Inert Doublet Model and its extension with an additional complex singlet is considered. The CP violation aspects are analysed in the simplified case, with one  SM-like Higgs doublet and a complex singlet.  
\end{abstract}


\PACS{12.60.Fr, 14.80.Ec, 14.80.Fd, 95.35.+d}
  
\section{Introduction}
A 125 GeV Higgs-like particle discovered at the LHC in 2012 has properties expected in the Standard Model (SM).   
However, as it is well known, the SM-like Higgs scenario appears in various models, among them in  models with two Higgs doublets (2HDM).   Models with more doublets, as well as with singlets, can be also considered, all leading to   $\rho=\frac{M_W^2}{M_Z^2 \cos^2 \theta_W} = 1$ at the tree-level, very close to the observed value. 

The SM-like Higgs scenario, as well as a stable scalar Dark Matter (DM), appears naturally within the  Inert Doublet Model  (IDM) -- a version of the Two Higgs Doublet Model with an exact discrete $D(Z_2)$  symmetry \cite{Deshpande:1977rw,Ma:2006km,Barbieri:2006dq}. This model is in agreement with current data, both from accelerator and astrophysical experiments \cite{Krawczyk:2013jta,Ilnicka:2015jba}. It has been applied to describe  the temperature evolution of the Universe in  $T^2$-approximation \cite{Ginzburg:2010wa}. There are possible sequences of phase transitions in the early Universe, that may lead to the current phase with a DM candidate in one, two or three steps. Using an effective potential approach with one-loop $T=0$  Weinberg-Coleman term and temperature-dependent effective potential at $T\not =0$ it was found, that the strong first-order phase transition needed for baryogenesis can be realized in the IDM; moreover,  a region with a good DM relic density and a strong first-order phase transition  exists \cite{Gil:2012ya}. 
However, the IDM cannot be a correct model for baryogenesis, because the CP violation occurs  in this model  as in the SM, i.e. only in the CKM matrix, which is well-known to be not sufficient. Therefore,  an extension of the IDM  is  desired. 

One of the option is adding one extra complex singlet of the scalar fields. We have studied this model, called the IDMS,  in Ref. \cite{Bonilla:2014xba}. 
We chose to have spontaneous CP violation in the model through a non-zero complex vacuum expectation value (VEV) of the singlet. 
There are three Higgs particles  with no definite CP properties; the lightest one among them can play a role of  a SM-like 125 GeV  Higgs particle.  On the other hand,  the lightest neutral scalar from the inert sector remains  a viable  DM candidate. Due to the extended Higgs sector, new interesting phenomena appear in the study of relic density, including new annihilation channels, interference between various Higgs contributions, and heavy Higgs resonance annihilation.  

Detailed analysis of CP properties of the  IDMS is rather complicated, so we have decided to consider first these properties in the simplified model, consisting of the SU(2) doublet and a complex singlet with non-zero VEV, which is a part of our IDMS framework. We found that to have the CP violation one non-zero cubic term in the singlet potential is needed.

In this work we present the IDM (section \ref{model}), together with its extension, the IDMS (section \ref{secIDMS}). We also discuss in details the differences between two models. The simplified case, with one  SM-like Higgs doublet and a complex singlet (SM+CS), is presented in section \ref{secSMCS}.  

\section{The IDM\label{model}}

Here we consider a particular version of 2HDM -- the Inert Doublet Model (IDM), which is very similar to the SM, as only  one SU(2) doublet, $\Phi_S$,
is involved in the Spontaneous Symmetry Breaking (SSB), and there exists only one SM-like Higgs particle.
The properties of the second doublet, $\Phi_D$, are quite different: it is not involved in the SSB and does not interact with fermions. It  contains four dark (inert) scalars, which have limited interactions with the SM particles and the lightest of them is stable, thus, if neutral, being a good candidate for Dark Matter.

The real content of the theory is determined by the symmetry properties of the Lagrangian, as well as of the vacuum state. 
We assume that the potential and the Yukawa interaction (only $\Phi_S$ interacts with fermions) are $Z_2$-symmetric with respect to the transformation $\Phi_S\to \Phi_S,\,\,\, \Phi_D\to -\Phi_D$, which we call  the $D$ symmetry.  Note, that this symmetry leads to the CP conservation in the model. The $D$-symmetric  potential has the following form:
\begin{displaymath}\begin{array}{c}
V=-\frac{1}{2}\left[m_{11}^2(\Phi_S^\dagger\Phi_S)\!+\! m_{22}^2(\Phi_D^\dagger\Phi_D)\right]+
\frac{\lambda_1}{2}(\Phi_S^\dagger\Phi_S)^2\! 
+\!\frac{\lambda_2}{2}(\Phi_D^\dagger\Phi_D)^2\\[2mm]+\!\lambda_3(\Phi_S^\dagger\Phi_S)(\Phi_D^\dagger\Phi_D)\!
\!+\!\lambda_4(\Phi_S^\dagger\Phi_D)(\Phi_D^\dagger\Phi_S) +\frac{\lambda_5}{2}\left[(\Phi_S^\dagger\Phi_D)^2\!
+\!(\Phi_D^\dagger\Phi_S)^2\right],
\end{array}\label{pot}\end{displaymath}
with all  parameters real. We take $\lambda_5<0$ without loss of generality \cite{Ginzburg:2010wa}. 

Various extrema 
can be realized in this potential. The possible vacuum expectation values (VEVs) are as follows:
\begin{equation}
\langle\Phi_S\rangle = \left( \begin{array}{c} 0 \\ \frac{1}{\sqrt{2}}v_S \\ \end{array} \right), \qquad   
\langle\Phi_D\rangle = \left( \begin{array}{c} u \\ \frac{1}{\sqrt{2}}v_D\\ \end{array} \right)
\quad v_S, v_D, u \in {R}. \label{dekomp_pol}
\end{equation}
Neutral extrema are realized for $u=0$. There are four types of neutral extrema (and also vacua) that have different symmetry properties:
\begin{itemize}
\item[(i)] The EWs case with $u = v_D = v_S = 0$  corresponds to the electroweak (EW) symmetry. 
\item[ (ii)] Mixed vacuum (M) with  $u = 0, v_S \not =0, v_D \not =0 $. There exist charged Higgs particles $H^\pm$, a pseudoscalar Higgs $A$ and two CP-even Higgses $h$ and $H$, either of them could be SM-like.
\item[(iii)] Inert vacuum (I$_1$) with  $u = v_D = 0, v_S\not =0$, $v_S^2=m_{11}^2/\lambda_1$,
conserves the $D$-parity and assures the existence of a stable scalar particle. 
\item[(iv)] Inertlike vacuum (I$_2$) with $u = v_S = 0, v_D\not =0$. This vacuum spontaneously violates the $D$ symmetry by $v_D \not = 0$,
$v_D^2=m_{22}^2/\lambda_2$. Fermions are massless.
\end{itemize}
If $u \not = 0$ then the charged vacuum (CB) is realized,  with the U(1)$_{\rm QED}$ symmetry breaking and the appearance of a massive photon. Such a case is not realized in the nature today.
Mixed and charged minima cannot exist for the same values of the parameters of the potential $V$. On the other hand, minima of the inert-type (I$_1$ or I$_2$) can overlap one another and with M and CB minima in the  $(\lambda_4, \lambda_5)$ parameter plane \cite{Krawczyk:2009fb}.

A stable vacuum (i.e. extremum with the lowest energy)  exists only if  positivity conditions hold
\begin{equation}
\la>0,\quad\lb>0,\quad\lc+\sqrt{\la\lb}>0,\quad\lczp+\sqrt{\la\lb}>0\,  , \label{stability}
\end{equation}
($\lczp=\lambda_3+\lambda_4+\lambda_5$) so that $$R= \frac{\lambda_{345}}{\sqrt{\lambda_1}\sqrt{\lambda_2}} > -1. $$ 
VEVs (\ref{dekomp_pol}) are related to the parameters of the potential through the extremum conditions, and so the various values of $v_S, v_D, u$ can be  represented  on  the 
phase diagram $(\mu_1,\mu_2)$, where
$\mu_1 =\frac{m_{11}^2}{\sqrt{\la}},\quad \mu_2=\frac{m_{22}^2}{\sqrt{\lb}}$.
Different regions of this parameter space correspond to the different vacua. 
There are three  regimes of parameter $R, \, R>1, 0<R<1, \,-1< R<0 $, which correspond to very different phase patterns.  
Note, that only for $R>1$ there is a  unique possibility of  coexistence of two inert-type minima, I$_1$ and I$_2$ \cite{Sokolowska:2011aa,Sokolowska:2011yi,Sokolowska:2011sb}.

In the Inert Doublet Model  I$_1$ is the  vacuum state. Here only one doublet ($\Phi_S$) is involved in the SSB and there is only one  SM-like Higgs boson $h$ -- we assume that its mass is equal to 125 GeV. The doublet $\Phi_D$ is inert (it has VEV $= 0$) and contains four scalars
 $H,\,A,
H^\pm$. Yukawa interactions are as in Model~I od 2HDM, since $\Phi_D$ does not interact with fermions. The $D$ symmetry is exact here and the lightest neutral scalar $H$ (or $A$) may play a role of DM. We take  DM $=H$ (so $\lambda_{4}+\lambda_5<0$).

Masses of the scalar particles read:
$$M_{h}^2=\lambda_1v^2= m_{11}^2\,,$$
$$ M_{H^\pm}^2= \frac{\lambda_3 v^2-m_{22}^2}{2}\,,
M_{A}^2=M_{H^\pm}^2+ \frac{\lambda_4-\lambda_5}{2}v^2\,, M_{H}^2=
M_{H^\pm}^2+ \frac{\lambda_4+\lambda_5}{2}v^2\,,
$$
with $v=246$ GeV. 

 Parameter $\lambda_1, m_{11}^2$ are fixed by the mass of 125 GeV Higgs particle, parameter $\lambda_{345}$ is related to  triple and quartic couplings between the SM-like Higgs $h$ and the DM candidate $H$, $\lambda_2$ gives the quartic DM self-couplings, while $\lambda_3$ describes the Higgs particle interaction with charged scalars. These parameters are  subject to various  theoretical and experimental constraints 
 (see e.g.~\cite{Barbieri:2006dq}, \cite{LopezHonorez:2006gr}\nocite{Tytgat:2007cv, Cao:2007rm, Gustafsson:2007pc, Agrawal:2008xz, Lundstrom:2008ai, Dolle:2009fn, Dolle:2009ft}--\cite{Arina:2009um},\cite{ Krawczyk:2009fb},\cite{Gustafsson:2009}\nocite{Gustafsson:2010, Honorez:2010re, LopezHonorez:2010tb, Kanemura:1993, Akeroyd:2000}--\cite{Swiezewska:2012}).

In particular, perturbative unitarity  leads to the following limits for self-couplings \cite{Swiezewska:2012}:
\begin{displaymath}
\lambda_{1,2}^{\textrm{max}} = 8.38, \quad \lambda_3 \in (-1.32, 16.53), \quad \lambda_{345} \in (-1.45,12.38).
\end{displaymath}
Above conditions, combined with requirement that I$_1$ is a global minimum \cite{Ginzburg:2010wa}, lead to~\cite{Swiezewska:2012}: 
\begin{displaymath}\label{m22bound}
m_{22}^2\lesssim 9\cdot10^4\g^2.
\end{displaymath}

  Also  EWPT strongly constrain  physics beyond SM. Important radiative corrections to gauge bosons propagators can be parametrized by the oblique parameters $S$ and $T$ \cite{Peskin:1990zt}. Values of these parameters are demanded to lie within $2\sigma$ ellipses in the $(S,T)$ plane, with the following central values~\cite{Nakamura:2010}: $S=0.03\pm0.09$, $T=0.07\pm0.08$, with correlation equal to 87\%. 
The LEP II analysis (reinterpretation of the MSSM analysis for the IDM ) excludes the region of masses where simultaneously: $M_{H} < 80$ GeV, $M_{A} < 100$ GeV and $M_A-M_H > 8$ GeV. For $M_A-M_H <8$ GeV the LEP I limit $M_{H} + M_{A} > M_Z$ applies \cite{Gustafsson:2009,Gustafsson:2010}. The LEP limit for $\m$ is: $\m\gtrsim 70 \, \g $ \cite{H+lep}.

\begin{figure}
\begin{center}
\includegraphics[width=.7\textwidth]{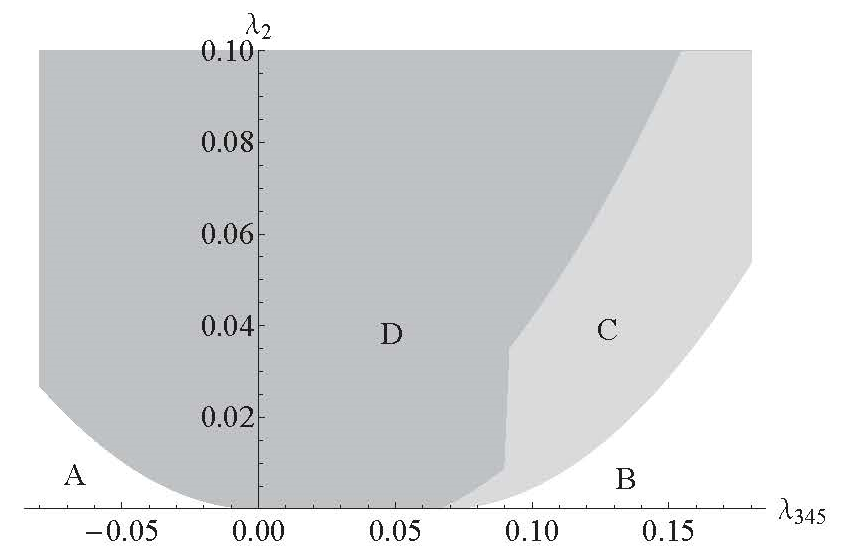}
\caption{Parameter space $(\lambda_{345}, \lambda_2)$ showing the coexistence of minima I$_1$ and I$_2$. Region A is excluded by positivity constraints. I$_1$ and I$_2$ coexist in regions B and C. In region B I$_2$ is is the global minimum, while I$_1$ is local, in region C the situation is reversed. In region D only minimum I$_1$ exists, from \cite{dokt-DS}.}
\label{l2}
\end{center}
\end{figure}

Note, that  direct measurements of  quartic self-coupling for dark scalars $\lambda_2$ is doubtful.  However,  there are some  indirect constraints for $\lambda_2$ that come from its relation to $\lambda_{345}$ through the vacuum stability constraints (\ref{stability}) and existence of I$_1$ vacuum \cite{Sokolowska:2011aa,Sokolowska:2011sb}, see Fig.(\ref{l2}).

\subsection{LHC Higgs data constraining parameters of the IDM}
The relevant LHC data on the discovered Higgs particle   constraining parameters of the IDM  are:
\begin{itemize}
\item Higgs mass $M_h = 125.09 \pm 0.24 $  GeV \cite{Aad:2015zhl},
\item Higgs total decay width $\Gamma_{\rm tot} = (4.2- 5.5)\times4.5$ MeV \cite{gamma},
\item Higgs total signal strength\footnote{Signal strength is often denoted by $\mu$.} $\mathcal{R} =1.09\pm 0.11$ \cite{comb},
\item Higgs decay into $\gamma \gamma$ signal strength $\mathcal{R}_{\gamma\gamma}=1.16^{+0.20}_{-0.18}$ \cite{comb},
\item Invisible decay branching ratio $\textrm{Br}(h \to {\rm inv}) <0.23-0.36$  \cite{invisible}.
\end{itemize}
Obviously, the Higgs boson in the IDM, $h$, has tree-level decays to the SM particles as in the SM, however decays $h \to \textrm{SM } \textrm{SM }$, which are absent at the tree level, may differ  slightly, e.g. $h \to gg, \gamma \gamma, Z \gamma$ proceeding via the loop couplings. These channels  cannot change the total decay width of $h$ considerably. Much more important  in this respect are  new tree-level decay channels, which are  absent  in the SM. In the IDM there are decays of $h$ to  inert particles, among them the $h \to HH$, called the  invisible decay.  The recent detailed analysis shows that the IDM is in agreement with all current LHC data \cite{Ilnicka:2015jba}.
   
Other  attempts undertaken at the LHC to constrain DM in processes independently from the Higgs sector, e.g. monojets or monophotons events,  are not leading to important limits for the IDM.  This is mainly because the IDM is a Higgs-portal DM model, where interaction between  DM and SM particles are dominated by exchange of   Higgs particle, with mass 125 GeV, while experimental analyses are based on effective coupling with heavy portal-particles. 

\subsubsection{Higgs Invisible decays}
The Higgs boson of the IDM, apart from the SM decay channels, has additional ones:  to a pair of dark particles, $AA$, $HH$ or $H^{\pm}H^{\mp}$.
  The decay width for the process $h\to HH$ reads (see e.g. Ref.~\cite{Krawczyk:2013jta})
$$
\Gamma(h\to HH)=\frac{\lambda_{345}^2 v^2}{32\pi M_h}\sqrt{1-\frac{4M_{H}^2}{M_h^2}},
$$
where $\lambda_{345}=\lambda_3+\lambda_4+\lambda_5$ is the coupling between the Higgs boson and a pair of DM particles.  For the decay $h\to AA$ the parameters $\lambda_{345}$ and $M_H$ have to be replaced by $\lambda_{345}^-=\lambda_3+\lambda_4-\lambda_5$ and $M_A$, respectively. 

The branching ratio of the Higgs boson decay to invisible particles is constrained by the LHC measurements, since the decay width depends on the mass of the product of the decay and its coupling to the Higgs boson.
 In the same way the measurement of the total Higgs decay width can be used. Below, for the sake of simplicity, we will assume that $A$ is too heavy for the $h\to AA$ process to be allowed, i.e. only $M_H <M_h/2$.
The allowed region on $\lambda_{345}$ and $M_H$ coming from the constraint on Br$(h\to \mathrm{invisible})<0.37$~\cite{invisible} and on the total width $\Gamma(h)<5.4\   \Gamma(h)^{\mathrm{SM}}$~\cite{gamma}  corresponds to  a small range around zero, $|\lambda_{345}|<0.05$, for masses of $ H$ below 62.5 GeV.

\subsubsection{Higgs decays to $\gamma\gamma$}
The results of  measurements  of Higgs decays to $\gamma\gamma$, mentioned above,  are close to the SM predictions, 
$\rgg=1$.  

Let us define $R_{\gamma\gamma}$ (see e.g. Refs.~\cite{Posch, Arhrib, Swiezewska:2012eh})
\be
\rgg:=\frac{\sigma(pp\to h\to \gamma\gamma)^{\textrm{IDM}}}{\sigma(pp\to h\to \gamma\gamma)^{\textrm  {SM}}}\approx\frac{\textrm{Br}(h\to\gamma\gamma)^{\textrm {IDM}}}{\textrm{Br}(h\to\gamma\gamma)^{\textrm {SM}}},
\ee
where the approximation of the narrow width of $h$ and the fact that the gluon fusion dominates in the Higgs production  were used, and therefore $\sigma({pp\to h})$  is the same in the IDM and the SM.

The $\textrm{Br}(h\to\gamma\gamma)^{\textrm {IDM}}$ depends both on the partial decay width of the Higgs boson to two photons, and on the total decay width of the Higgs particle.
Note that, to a good approximation,  only the invisible channels modify the total decay width of the Higgs boson with respect to the SM. 
Note that once the invisible channels are kinematically allowed, they dominate over the SM channels, so in general they tend to suppress $\rg$.

In the SM the $h\to \gamma\gamma$ decay is induced by a $W^{\pm}$ boson loop and  a fermionic loops (the top quark dominates).
$\Gamma(h\to \gamma\gamma)$ in the IDM differs from the one computed in the SM, because of the charged scalar, $H^{\pm}$, 
 which gives an extra contribution to the process. This contribution can interfere either constructively or destructively with the SM part,
so Br$(h\to \gamma\gamma)^{\textrm {IDM}}$ can be enhanced or suppressed with respect to the SM.


Let us first analyse the consequences of enhanced signal strength (we follow Ref.~\cite{Swiezewska:2012eh}). 
It was found that for $M_H<M_h/2\approx \, 62.5 \, \g$ the diphoton signal strength is always suppressed with respect to the SM, due to enhancement of the total decay width of $h$. This means that if enhancement of the signal strength is observed, DM with mass below 62.5 GeV is excluded.
Note, that if $\rgg>1.2$, then only fairly light charged scalar  and $H$ (as $M_H<\m$)  are allowed, with $\m, M_H \lesssim 154 \,\g$.

The case where we allow $\rgg$ to go below 1 will be analysed in the next section, and combined with the DM  measurements.

\subsection{Relic density constraints}
The IDM provides a good DM candidate ($H$) in agreement with the data on relic density $\Omega_{DM} h^2$, $\Omega_{DM} h^2=0.1199 \pm 0.0027$ \cite{Ade:2013zuv}, in three regions of $M_H$, see eg. 
\cite{Barbieri:2006dq, Sokolowska:2011aa}, \cite{Sokolowska:2011sb}\nocite{LopezHonorez:2006gr, Tytgat:2007cv,
Cao:2007rm}--\cite{Gustafsson:2007pc}, \cite{Dolle:2009fn}\nocite{Dolle:2009ft}--\cite{Arina:2009um},  \cite{Honorez:2010re,LopezHonorez:2010tb}:
\begin{itemize}
\item light DM particles with masses below $10 \textrm{ GeV}$. Other dark particles are much heavier than DM and so this scenario mimics the behavior of the singlet DM model. As the Higgs-DM coupling needed to obtain proper relic density in this region is relatively large, this region of masses is excluded by combined results of relic density measurements and LHC data on Higgs invisible decays \cite{Krawczyk:2013jta}.
\item  medium mass regime of $50-150 \textrm{ GeV}$ with two distinctive regions: with and without coannihilation of $H$ with the  neutral $D$-odd particle~$A$. Coannhilation channels are present, if the mass difference(s) between $H$ and other scalar particle(s) are small. For both cases, i.e. with  and without  coannihilation of $H$ and $A$, 
the allowed region of $\lambda_{345}$ is symmetric around zero for masses $M_H \lesssim 72$~GeV. Usually, very small values of $\lambda_{345}$ are excluded due to a non efficient DM annihilation. If coannhilation channels are open, allowed values of $\lambda_{345}$ are smaller than if coannhilation channels are closed, as the process $HH \to h \to \bar{f}{f}$ with the cross-section $\sigma \sim \lambda_{345}^2$ does not have to be that efficient to provide the proper relic density value. As  mass increases, the region of proper relic density shifts towards the negative values of $\lambda_{345}$, which is due to opening of the annihilation channels into the gauge bosons final state and interference of processes $HH \to h \to VV$ and $HH \to VV$. For the IDM, it is possible to obtain correct relic density for masses up to 150 GeV. However, larger values of masses, which correspond to larger values of $\lambda_{345}$, are excluded by direct detection limits coming from LUX experiment.
\item heavy DM of mass larger than $525 \textrm{ GeV}$. In this region all dark particles have almost degenerate masses and coannihilation processes between all dark particles are crucial for the agreement with the measured  DM relic density.
\end{itemize}
 The relic density data can be used to constrain   $\lambda_{345}$ coupling for  chosen values of masses of $H$ and other scalars \cite{Dolle:2009fn,LopezHonorez:2006gr}. The same coupling, related to Higgs-charged scalar coupling $\lambda_3$, influences the values of $\rgg$ \cite{Krawczyk:2013jta}. It was found that for DM with masses $M_H<M_W$ it is not possible to have an enhancement in the $h \to \gamma \gamma$ signal, and, at the same time, to be in agreement with relic density constraints. In this region $\rgg$ is always below the SM value, unless the DM relic density is too small. For heavy DM candidate, as the DM influence on the Higgs sector is reduced, $\rgg \approx 1$.

\section{The IDMS \label{secIDMS}}

The IDM provides a viable DM candidate in a wide range of DM masses. However, it lacks one important ingredient needed to explain the matter-antimatter asymmetry of the Universe, namely an additional, with respect to the SM, source of CP violation. In this section we present a model which can provide this desired addition, the IDMS, i.e. the IDM plus a complex singlet \cite{Bonilla:2014xba}. The IDMS is a $Z_2$-symmetric model that contains a SM-like Higgs doublet $\Phi_1$, as well as an inert scalar doublet $\Phi_2$,
which has VEV=0 and is odd under a $Z_2$ symmetry, and the neutral complex singlet $\chi$, with hypercharge $Y=0$  and a non-zero complex VEV.\footnote{Here we use notation $\Phi_1$ and $\Phi_2$ for the SM-like doublet and the inert doublet, respectively.}

There are different choices of symmetries in this kind of model that can lead to different phenomenology. Here, we shall assume that the symmetry assignment is as follows: 
\begin{equation}
Z_2\;:\; \Phi_1 \to \Phi_1, \;  \Phi_2 \to - \Phi_2, \; \textrm{SM fields} \to  \textrm{SM fields}, \; \chi \to \chi. \label{IDMSz2}
\end{equation}
This choice results in an inert sector just like in the IDM, while the Higgs sector consists of both $\Phi_1$ and $\chi$.  In our model only $Z_2$-even fields $\Phi_1$ and $\chi$ acquire vacuum expectation values, leading to the following field decomposition around the vacuum state $(v,0,w e^{i\xi})$, where  $v,w, \xi \in \bf{R}$:
\begin{eqnarray}
& \Phi_{1} = \left( \begin{array}{c} \phi^+_1 \\ \frac{1}{\sqrt{2}} \left( v + \phi_1 + i \phi_6 \right)\\ 
\end{array} \right), \qquad \Phi_{2} = \left( \begin{array}{c}\phi^+_2 \\ \frac{1}{\sqrt{2}} \left(  \phi_4 + 
i \phi_5 \right)\\ \end{array} \right), \label{IDMSdec1}&\\[2mm]
& \chi = \frac{1}{\sqrt{2}} (w e^{i \xi} + \phi_2 + i \phi_3). \label{IDMSdec2}&
\end{eqnarray}
For simplicity  we will consider the constrained potential, the cIDMS \cite{Bonilla:2014xba}:
\begin{eqnarray}
  &
    \begin{array}{c}
V = -\frac{1}{2}\left[{m_{11}^2} \Phi_1^\dagger\Phi_1 + {m_{22}^2} \Phi_2^\dagger\Phi_2 \right] 
+ \frac{1}{2}\left[\lambda_1 \left(\Phi_1^\dagger\Phi_1\right)^2 
+ \lambda_2 \left(\Phi_2^\dagger\Phi_2 \right)^2\right]\\[6mm]
+  \lambda_3 \left(\Phi_1^\dagger\Phi_1 \right) \left(\Phi_2^\dagger\Phi_2\right) + \lambda_4 
\left(\Phi_1^\dagger\Phi_2\right) \left(\Phi_2^\dagger\Phi_1\right) +\frac{\lambda_5}{2}\left[\left(\Phi_1^\dagger\Phi_2\right)^2\!
+\!\left(\Phi_2^\dagger\Phi_1\right)^2\right] \\[3mm]
-\frac{m_3^2}{2} \chi^* \chi + \lambda_{s1} (\chi^*\chi)^2 + \Lambda_1(\Phi_1^\dagger\Phi_1)(\chi^* \chi)\\[2mm]
 -\frac{m_4^2}{2} (\chi^{*2} + \chi^2) + \kappa_2 (\chi^3 + \chi^{*3}) + \kappa_3 [ \chi(\chi^*\chi) + \chi^*(\chi^*\chi)].
    \end{array}&
\label{potIDM1S}
\end{eqnarray}
Here, we have imposed a global $U(1)$ symmetry: 
\begin{equation}
U(1): \;\; \Phi_1 \to \Phi_1,\, \Phi_2 \to \Phi_2, \, \chi \to e^{i\alpha} \chi \label{IDMSu1def}.
\end{equation}
and kept only $U(1)$-symmetric and $U(1)$-soft-breaking terms of the most general IDMS potential \cite{Bonilla:2014xba}.


Due to the imposed $Z_2$-symmetry, the neutral scalar sector is divided into two separate sectors:  the $Z_2$-even Higgs sector, and the $Z_2$-odd inert sector. The inert sector, $H,A$ and $H^\pm$, is just like in the IDM, with $H$ chosen to be the lightest inert particle, i.e. a DM candidate. Mass formulas for the inert particles from the IDM still hold, and they depend only on $m_{22}^2$ and  $\lambda_{3,4,5}$. Note, that $\chi$ does not influence the inert sector in the cIDMS.

The Higgs sector is extended with respect to the IDM by addition of the singlet $\chi$, which results in three Higgs particles $h_1,h_2,h_3$. If $\Lambda_1, w, \sin \xi \not =0$ then there is mixing between states of different CP properties. Relation between physical states $h_{1,2,3}$ and the base states with defined CP $\phi_{1,2,3}$ is given by:
\begin{eqnarray}
 & \left( \begin{array}{c} h_1\\ h_2\\ h_3\\ \end{array} \right) = R \left( \begin{array}{c} 
 \phi_1\\ \phi_2\\ \phi_3\\ \end{array} \right), & \label{neutr_diag}
\end{eqnarray}
where $R$ is a 3$\times$3 rotation matrix, that depends on three mixing angles (here and below $c_i = \cos \alpha_i, s_i = \sin \alpha_i$):
\begin{equation}
R = R_1 R_2 R_3 = \left(
\begin{array}{ccc}
c_1 c_2 & c_3 s_1 - c_1 s_2 s_3 & c_1 c_3 s_2 + s_1 s_3\\
-c_2 s_1 & c_1 c_3 + s_1 s_2 s_3 & -c_3 s_1 s_2 + c_1 s_3\\
-s_2 &  -c_2 s_3 & c_2 c_3
\end{array} \right).\label{rotfull}
\end{equation}
The rotation matrix and its inverse give us two important relations:
\begin{equation}
h_1 = c_1 c_2 \phi_1 + (c_3 s_1 - c_1 s_2 s_3) \phi_2 + (c_1 c_3 s_2 + s_1 s_3) \phi_3,
\end{equation}
\begin{equation}
\phi_1 = c_1 c_2 h_1 - c_2 s_1 h_2 - s_2 h_3, \label{phi1DM}
\end{equation}
which describe the composition of the SM-like Higgs boson $h_1$, in terms of real components  $\phi_1$ and  $\phi_2$, 
which provide a CP-even part, as well as the $\phi_3$ component -- CP-odd one. Equivalently, one can look at it as the modification of 
the real component of the SM-like Higgs doublet $\Phi_1$ from the cIDMS with respect to the SM and the IDM. Especially important is the first element both in $R$ and $R^{-1}$ equal to:
\begin{equation}
R_{11} = R^{-1}_{11} = c_1 c_2. \label{r11}
\end{equation}
This matrix element gives the relative modification of the interaction of the Higgs boson ($h_1$) with respect to the IDM, and is important both in the LHC analysis, and in the DM studies.

\subsection{Constraints}
The parameter space of the cIDMS is constrained by current theoretical and experimental results. We follow the limits that we use for the IDM, described in the previous section, and we ascertain that the extended scalar sector is not violating any of those constraints. In our analysis we have checked the agreement with LEP results on invisible decays of $W^\pm, Z$ and lower mass limit of the charged scalar; perturbativity conditions, which constrain mass splittings of inert particles in the heavy mass regime; EWPT limits, which test the influence of both inert scalars and additional Higgs particles. Furthermore, we shall demand that the lightest Higgs particle, $h_1$ is a SM-like Higgs with mass $M_{h_1} = 125$ GeV, and that it is in agreement with the latest LHC results. Finally, we will check that $H$ is a good DM candidate, with proper relic density and within direct detection limits.

\subsection{LHC Phenomenology}
\begin{figure}
\begin{center}
\includegraphics[width=0.45\textwidth]{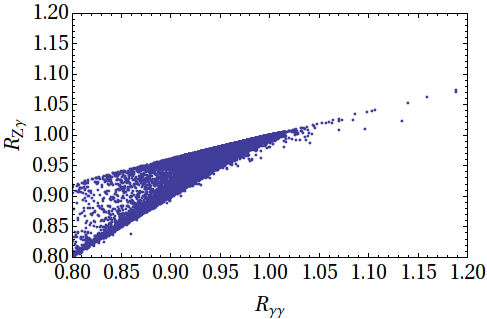}  \includegraphics[width=0.45\textwidth]{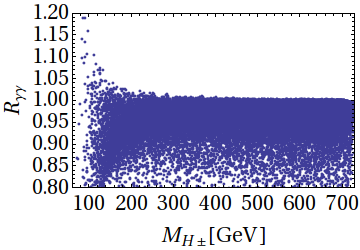}
\caption{Left: Correlation between $\mathcal{R}_{\gamma\gamma}$ and $\mathcal{R}_{Z\gamma}$. Right: Correlation between $\mathcal{R}_{\gamma\gamma}$ and $M_{H^\pm}$. Plots taken from \cite{Bonilla:2014xba}.}\label{Fa1}
\end{center}
\end{figure}

Latest LHC results provide further constraints on the parameters of the model. We consider  the signal strengths $\mathcal{R}_{gg}, \, \mathcal{R}_{\gamma \gamma}, \,\mathcal{R}_{Z\gamma},\, \mathcal{R}_{VV}$.
According to (\ref{phi1DM}) the couplings of the lightest Higgs particle ($h_1$) with 
vector bosons and top quark get modified, as compared with the SM, only by a factor $R_{11}$ (eq. (\ref{r11})). The one-loop coupling of $h_1$ to photons receives contributions mainly
from the W boson and top quark, as well as the charged scalar $H^\pm$ from the inert sector. Thus, we can write the relevant $h_1$ decay widths as follows\footnote{See Appendix A in \cite{Bonilla:2014xba}  and references therein for more details.}:
\begin{equation*}
    \begin{array}{c}
       \Gamma(h_1 \to XX )= R_{11}^2 \Gamma(\phi_{SM} \to XX), \; XX = gg, VV^{*}.\\[1mm]
 \Gamma (h_1\to \gamma \gamma) = R_{11}^2  |1+ \eta_{\gamma \gamma}|^2\Gamma (\phi_{SM} \to \gamma \gamma),
\eta_{\gamma \gamma}= \frac{ g_{h_1 H^+ H^-} v } { 2 R_{11} M^2_{H^{\pm}}}\left(\frac{A_{H^\pm}}{A^{SM}_W+ A^{SM}_t}\right) ,
\end{array}
\end{equation*}
and analogous formula for  $\Gamma(h_1\to Z\gamma)$.
Note, that the triple coupling $\lambda_{h_1 H^+ H^-}$ is given by $g_{h_1 H^+ H^-}=v \lambda_{3} R_{11}$,
meaning it is also modified with respect to the IDM by a factor of $R_{11}$. Finally, the signal strengths 
can be written as follows,
\begin{eqnarray*}
&\mathcal{R}_{ZZ}= R_{11}^2\zeta^{-1}, \ \ \mathcal{R}_{\gamma\gamma}= R_{11}^2 |1+ \eta_{\gamma\gamma}|^2\zeta^{-1},
\mathcal{R}_{Z\gamma}=R_{11}^2 |1+\eta_{Z\gamma}|^{2}\zeta^{-1},
\end{eqnarray*}
where $\zeta$ is defined as
\begin{equation*}
\zeta\equiv1+\frac{\Gamma_{\rm inv}}{R_{11}^2\Gamma_{\rm tot}^{SM}},
\end{equation*}
with $\Gamma_{\rm inv}$ corresponding to invisible decays of $h_1$ into inert particles (if $M_{H,A} < M_{h_1}/2$), and $\Gamma_{\rm tot}^{SM}$ being the total decay width of the SM Higgs boson. Notice that $\mathcal{R}_{ZZ} \leq 1$, while both $\mathcal{R}_{\gamma\gamma}$ and $\mathcal{R}_{Z\gamma}$ can exceed 1. If $\mathcal{R}_{\gamma\gamma}<1$ then both $\mathcal{R}_{Z\gamma}$ and $\mathcal{R}_{ZZ}$ are correlated 
with $\mathcal{R}_{\gamma\gamma}$, $\mathcal{R}_{\gamma\gamma}\sim\mathcal{R}_{Z\gamma}$ and $\mathcal{R}_{\gamma\gamma}\sim\mathcal{R}_{ZZ}$ (Fig. \ref{Fa1}).
 There is a possibility of similar enhancement of both $\mathcal{R}_{\gamma\gamma}$ and $\mathcal{R}_{Z\gamma}$. This is in agreement 
with the IDM, where a correlation between enhancement in $\gamma\gamma$ and $Z\gamma$ channels exists \cite{Swiezewska:2012eh}. 

For smaller values of $M_{H^\pm}$ there is a possibility of enhancement of both $\mathcal{R}_{\gamma\gamma}$ (Fig. \ref{Fa1}) and $\mathcal{R}_{Z\gamma}$. For heavier $M_{H^\pm}$ the maximum values tend to the SM value, with a possible deviation up to 20 \%. This is again a result similar to the one from the IDM, 
where significant enhancement, e.g. $\mathcal{R}_{\gamma\gamma} = 1.2$,
was possible only if $M_{H^\pm} \lesssim 150\, \g$, and for heavier masses $\mathcal{R}_{\gamma\gamma} \to 1$ \cite{Swiezewska:2012eh}.


\subsection{DM Phenomenology}

The cIDMS can be treated as an extension of the IDM, and certain properties of DM sector are kept here. As there is no direct coupling between the inert doublet $\Phi_2$ and the singlet $\chi$, and the only interaction is through mixing of $\chi$ with $\Phi_1$, the 
inert particles' interaction with gauge bosons is like in the IDM, while the inert scalars-Higgs boson
interaction changes. The IDM Higgs particle $h$ corresponds  in our case to $\phi_1$, so $h \to \phi_1$, where $\phi_1$ in terms of physical fields is given by (\ref{phi1DM}). In our analysis we are focusing on the SM-like scenario, which corresponds to $c_1c_2 \approx 1$. This would naively suggest that for all purposes we could keep $h \approx h_1$, and neglect the contribution from additional Higgs particles. However, our study shows that even if $c_1 c_2 \approx 1$ there are significant differences with respect to the IDM in relic density values coming from additional annihilation channels, interference and resonance effects due to the extended Higgs sector.

\subsubsection{Benchmark Points}
The analysis of DM properties was done for a couple of benchmarks\footnote{Additional benchmarks are discussed in \cite{Bonilla:2014xba}.} chosen in agreement with constraints from LHC/LEP:
\begin{eqnarray*}
&& \textrm{\textbf{A1}: } M_{h_1} = 124.83 \g, \; M_{h_2} = 194.46 \g, \; M_{h_3} = 239.99 \g, \\
&& \textrm{\textbf{A4}: } M_{h_1} = 125.36 \g, \; M_{h_2} = 149.89\g, \; M_{h_3} = 473.95 \g. 
\end{eqnarray*}
The above values were chosen to illustrate different possible scenarios: in \textbf{A1} all Higgs particles are relatively light, but only $h_1$ is lighter than $2M_W$; 
for \textbf{A4} there are two Higgs particles that have mass below $2M_W$: $h_1$ (the SM-like Higgs) and $h_2$. We treat $2M_W$ as the distinguishing value as two Higgs particles with $M_{h_i} < 2 M_{W}$ influence the DM phenomenology by introducing another resonance region in the medium DM  mass regime.

\subsubsection{Low, Medium and Heavy Mass Regions}

\begin{figure}
  \centering
\hspace*{-0.3cm}\includegraphics[width=0.8\textwidth]{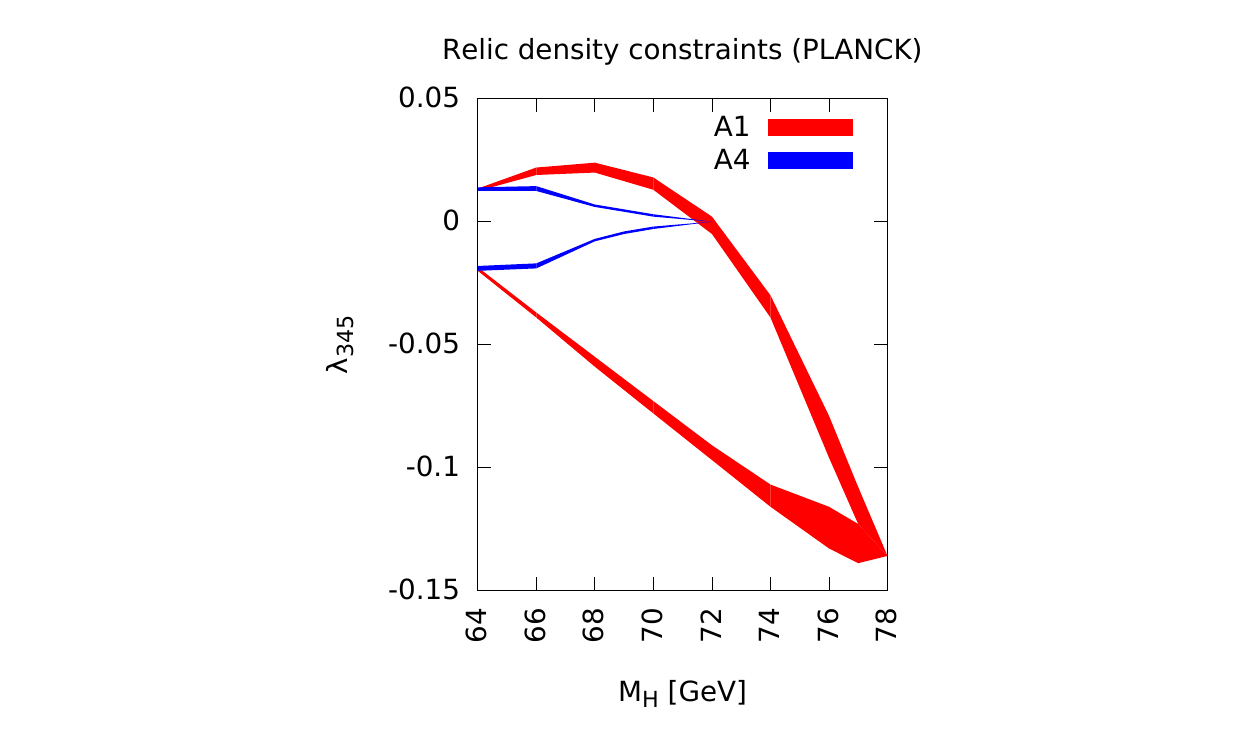}  \includegraphics[width=0.7\textwidth]{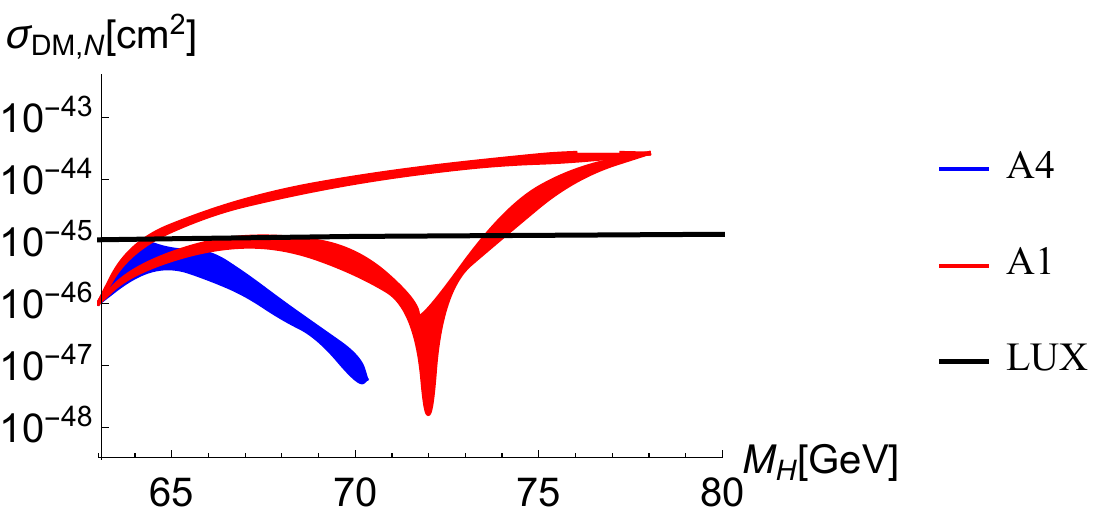}
      
\vspace{-5pt}
  \caption{Up: Relic density constraints on the mass of the DM candidate ($M_H$) and its coupling to SM Higgs boson. Red and blue regions corresponds to relic density in agreement with Planck measurements for  benchmark A1 and A4, respectively. Down: DM-nucleus scattering cross-section for benchmarks A1 (red) and A4 (blue), with respect to LUX limits. \label{mid}  }
\end{figure}

The cIDMS can provide a good DM candidate, with relic density in agreement with Planck results, in three regions of masses:\\[1mm]
$\bullet$ Light DM mass: $53\, \g \lesssim M_H \lesssim M_{h_1}/2$ with $M_A, M_{H^\pm} \gtrsim M_H + 50 \,\g$. Here, we observe no significant differences between benchmarks, because the main DM annihilation channel is the near-resonance $HH\to h_1 \to b \bar{b}$, thus making annihilation through heavier $h_{2,3}$ negligible. As in the IDM, the value of coupling that is in agreement with direct detection limits, LHC results and relic density measurements is very small, $\lambda_{345} \sim 0.002$ \cite{Bonilla:2014xba}.\\[1mm]
$\bullet$ Medium DM mass: $M_{h_1}/2 \lesssim M_H \lesssim M_W$ with $M_A, M_{H^\pm} \gtrsim M_H + 50 \,\g$. Figure \ref{mid} shows the relation between parameter $\lambda_{345}$ (which is closely related to the DM-Higgs coupling, $g_{HHh_1} = c_1 c_2 \lambda_{345}$, with $c_1 c_2 \approx 1$ for all  SM-like scenarios) and DM mass $M_H$ for benchmark A1 and A4. Benchmark A1, where both additional Higgs particles 
are heavier than $2M_W$, follows the well-known behaviour of the IDM.  However, the presence of these additional states is non-negligible, as the annihilation of DM particles is enhanced and therefore the relic density for a given mass is lower 
with respect to DM candidate from the IDM \cite{Bonilla:2014xba}.


A significant difference with respect to the IDM is visible if one of the extra Higgs bosons is lighter than $2M_W$, 
which is the case for benchmark A4. 
The annihilation rate is enhanced and dominated by the Higgs-type resonance
exchange through $h_2$ (note the symmetric distribution around $\lambda_{345} = 0$), in contrast 
to the previously discussed cases \cite{Bonilla:2014xba}.\\[1mm]
$\bullet$  Heavy DM mass: $M_H \gtrsim 525\, \g$ with $M_A = M_{H^\pm} = M_H + 1 \, \g$. Again, we observe no significant differences between benchmarks, and this region reproduces DM phenomenology from the IDM. This is because the heavy DM candidate annihilates mostly through gauge (co)annihilation channels, therefore the extended scalar sector has minimal influence on the results.

\subsubsection{Direct Detection Limits}
The cIDMS, as other scalar DM models, can be strongly constrained by results of direct detection experiments.
The current strongest limits come from LUX experiment \cite{Akerib:2013tjd}. As is the case with many other scalar DM models, the constraints are not strong enough for heavy DM candidate, as the sensitivity of direct detection experiments reduces significantly with increasing DM mass. The near resonance region $53 \; \g \lesssim M_H \lesssim 63 \; \g$ also escapes detection, due to very small value of Higgs-DM coupling\footnote{Recall that the DM scattering off nuclei is mediated by the Higgs particles, $h_1,h_2,h_3$, therefore the strength of this scattering will directly depend on the value of DM-Higgs couplings.}.

Fig. \ref{mid} presents the value of DM-nucleus scattering cross-section, $\sigma_{DM,N}$, for medium DM mass region, along with the latest LUX limit. Red regions denote benchmark A1, while blue regions correspond to benchmark A4. The difference between those two groups is clear. In case of benchmark A4, the coupling 
is usually much smaller than in case A1, therefore the resulting cross-section will be also 
smaller, falling well below the current experimental limits. 


\subsection{Comparison with the IDM}

Both the IDM, and the cIDMS, can provide a good DM candidate, which is in agreement with the current experimental results. In both models, it is not possible to fulfil LHC constraints for the Higgs invisible decay branching ratio and relic density measurements at the same time if $M_{H} \lesssim 53$ GeV. 
Masses $53 \; \g \lesssim M_H \lesssim 63 \; \g$ correspond to the resonance region of enhanced annihilation with very 
small coupling $\lambda_{345}$. It is important to stress that, while DM phenomenology of the cIDMS does not depend on the chosen 
benchmark point (A1-A4), there is a difference when it comes to the LHC observables. Values of $\rgg$ 
for benchmark A4 are smaller than in all other cases -- and smaller than the IDM -- being always below 1.

For heavier DM mass, the extended scalar sector significantly changes the annihilation rate of 
DM particles. Our studies show that the annihilation cross-section is enhanced with respect to the 
IDM and therefore relic density in the cIDMS is usually lower than for the corresponding point in the 
IDM. Especially important is the possibility of having a second resonance region, that will mimic the low DM mass behaviour, if the mass of one of additional Higgs particles is smaller than $2M_W$. 
Corresponding DM-Higgs couplings, and thus the 
resulting $\sigma_{DM,N}$, constrained by results of direct detection experiments, is 
much smaller for A4 than for other benchmark points.
 
Heavy DM mass region resembles the IDM very closely.
In the heavy mass region all inert particles are heavier than the particles from the SM sector and 
the impact on the Higgs phenomenology can be minimal. For example, this is the region where $\rgg$ 
is the closest to the SM value. 

The possibility of having  CP violation is a significant modification of our model with respect to the IDM. It changes not only interaction in the Higgs sector, by allowing three states $h_1$, $h_2$ and $h_3$ to have non-zero coupling to $VV$ pairs, but also influences the DM sector, which itself -- by construction -- is CP conserving. CP violation in the Higgs sector changes the annihilation channels; for example,  channels like $HH\to h_i \to Z h_j$ can appear and significantly change the relic density value if DM particle is heavy enough.

\section{The SM+CS: The SM plus a complex singlet \label{secSMCS}}
As stated  before, the CP violation in the SM is insufficient to explain the baryon asymmetry in the Universe and therefore an additional source of CP-violation is needed. The simplest possibility is  through the extension of SM with a complex singlet \cite{Branco:2003rt, Bento:1991ez}. We consider such a scenario according to which the SM-like  Higgs particle comes mostly from the SM-like SU(2) doublet, with a small modification coming from the singlet. Note, that this is a part of the IDMS responsible for the CP violation.

The scalar potential of the model, with assumption on U(1) symmetry as in cIDMS (\ref{IDMSu1def}), can be written as:
$$
V = -\frac{1}{2}{m_{11}^2} \Phi_1^\dagger\Phi_1+ \frac{1}{2}\lambda_1 \left(\Phi_1^\dagger\Phi_1\right)^2 
-\frac{m_s^2}{2} \chi^* \chi + \lambda_{s1} (\chi^*\chi)^2 + \Lambda_1(\Phi_1^\dagger\Phi_1)(\chi^* \chi)
$$
\begin{equation}
 -\frac{m_4^2}{2} (\chi^{*2} + \chi^2)+ \kappa_2 (\chi^3 + \chi^{*3})  + \kappa_3 [ \chi(\chi^*\chi) + \chi^*(\chi^*\chi)].
\label{potchi}
\end{equation}
 $\Phi_1$ and $\chi$ fields acquire vacuum expectation values $v$ and $w e^{i\xi}$, respectively, where  $v,w,\xi\in \bf{R}$. We shall use the following field decomposition around the vacuum state:
\begin{eqnarray}
\Phi_{1} = \left( \begin{array}{c} \phi^+_1 \\ \frac{1}{\sqrt{2}} \left( v + \phi_1 + i \phi_4 \right)\\ \end{array} \right), \qquad
\chi = \frac{1}{\sqrt{2}} (w e^{i \xi} + \phi_2 + i \phi_3). \label{dec_singlet}
\end{eqnarray}

We assume that all parameters of $V$ (\ref{potchi}) are real. There are three quadratic parameters, six dimensionless quartic parameters and five dimensionful parameters $\kappa_{i},\;i=1,2,3,4,5$. The linear term $\kappa_1$ can be removed by a translation of the singlet field, and therefore neglected. 

In order to have a stable minimum, the parameters of the potential have to
satisfy the positivity conditions. Particularly, the potential should be bounded from below, i.e. should not go to negative
infinity for large field values. As this behavior is dominated by the quartic terms, the cubic terms will not play a role here. Thus, the following conditions will apply to a variety of models that will differ only by their cubic interactions. The positivity conditions read:
\begin{equation}
\begin{array}{l}
 \lambda_1,\lambda_{s1} > 0,\qquad
 {\bar{\lambda}_{1S}} = \Lambda_1  + \sqrt{2 \lambda_1 \lambda_{s1}} >
 0.\\[3mm]
 \end{array} \label{pos}
\end{equation}

\subsection{Extremum  conditions}
It is useful to decompose complex fields into two real fields $S_1,S_2$ with respective VEVs $w_1,w_2$:
\begin{equation}
\begin{array}{c}
\chi=({S_1+i S_2})/{\sqrt{2}}, \,	\chi^*=({S_1-i S_2})/{\sqrt{2}},\\[1mm]
\langle \chi\rangle =w e^{i\xi}=\underbrace{w\cos\xi}_{w_1}+ i \underbrace{w\sin\xi}_{w_2}; \qquad w_1^2+w_2^2=w^2.
\end{array}
\label{VEV}
\end{equation}

\begin{figure}
\centering
  \includegraphics[width=.65\textwidth]{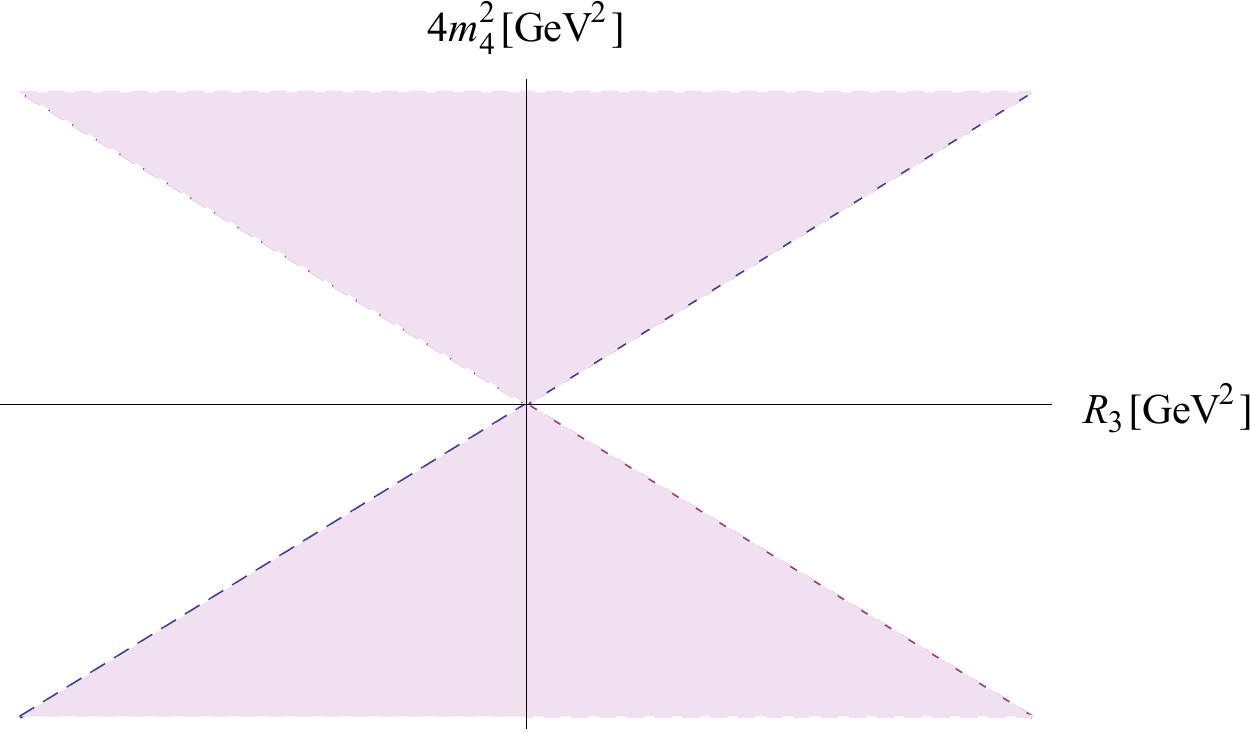}
\caption{  $(R_3,\, m_4^2)$ for $R_2=0$, $\cos \xi \in (-1,1)$. }
\label{Fig:F4H}
\end{figure}

For potential (\ref{potchi}), we got the following extremum  conditions:
	\begin{equation}
	-m_{11}^2+\lambda v^2+ \Lambda_1 w^2 =0, \label{min1}
	\end{equation}
	$$
	w_1 (-m_{s}^2-2m_4^2+v^2 \Lambda_1 + 2 w^2\lambda_{s_1}) 
	+ 3\sqrt{2}(w_1^2 - w_2^2)  \kappa_2 
	$$
	\begin{equation}
	+ \sqrt{2}(3 w_1^2 + w_2^2)  \kappa_3=0, \label{min2}
	\end{equation}
	\begin{equation}
	w_2(-m_{s}^2+2m_4^2+ v^2 \Lambda_1 + 2 w^2 \lambda_{s_1} + 2 \sqrt{2}  w_1 (- 3\kappa_2 + \kappa_3))=0.
	\label{min3} \end{equation}
We will keep $w_1$ and $w_2$ non-zero, noticing that only when  $ w_2\neq0$, there is a non-zero phase of $\chi$. 
Performing   the subtraction of equation (\ref{min2}) from equation (\ref{min3}) we obtain an  important relation between parameters:
\begin{equation}
-4 m_4^2 \cos\xi +3 R_2 (1+2\cos2\xi)+R_3=0,
\label{CP}
\end{equation}
where  $R_2=\sqrt{2}w{^2}\kappa_2$ and $R_3=\sqrt{2}w^2\kappa_3$. This relation  describes the region of parameters in  which CP violation may  occur.  The CP violation in the SM+CS model,  due to  non-zero phase of the complex singlet,  may be realized in wide regions  determined  by   quadratic and cubic parameters from $V$.
\begin{figure}
\begin{center}
\includegraphics[width=.65\textwidth]{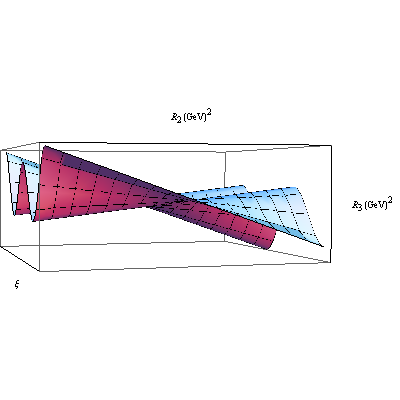}
\caption[CP violation]{ $(R_2,R_3,\xi)$ for  $m_4^2=500\; \g^2$.}
\label{Fig:F1H}
\end{center}
\end{figure}
Fig. \ref{Fig:F1H} and Fig.~\ref{Fig:F4H} show regions of parameter space where this relation holds. Fig.~\ref{Fig:F1H} shows the correlation between $R_2$, $R_3$ and $\xi$  for $m_4^2=500\;\g^2$,
while Fig.~\ref{Fig:F4H} presents the $4m_4^2$ against the $R_3$ for $R_2$=0. 

One can conclude that for the existence of a viable CP violation in the model it is necessary to have at least two  non-vanishing variables among  $R_2,\, R_3,\, m_4^2$.  

\section{Conclusions} 

The Inert Doublet Model is simple, yet very powerful, extension of the SM scalar sector. It can provide a good DM candidate, and at the same time accommodate the SM-like Higgs signal. The IDM is in agreement with all current experimental data. However, the IDM is missing one crucial ingredient to explain the baryon asymmetry of the Universe, i.e. the additional source of CP violation. Further extensions of the IDM, e.g. by a complex singlet, can solve this problem, by introducing explicit or spontaneous CP violation in the scalar sector.  Furthermore, the extended scalar sector can influence the DM phenomenology, including not only changes in DM annihilation scenarios, but also changing prospects of DM detection, either by dedicated DM direct detection experiments, or by the LHC.

\section{Acknowledgments}
MK and ND would like to thank organizers of this and all previous Schools one of us attended.  Presented  results were obtained in fruitful collaborations with B. \'{S}wie\.zewska, P. Swaczyna, J.~L.~Diaz-Cruz and C. Bonilla. 

The work  was partially supported by the grant NCN OPUS 2012/05/B/ ST2/03306 (2012-2016).

\end{document}